# A ROBUST MECHANISM FOR DEFENDING DISTRIBUTED DENIAL OF SERVICE ATTACKS ON WEB SERVERS

Jaydip Sen

Innovation Labs, Tata Consultancy Services Ltd.,
Bengal Intelligent Park, Salt Lake Electronic Complex, Kolkata, INDIA
`Jaydip.Sen@tcs.com`

## ABSTRACT

*Distributed Denial of Service (DDoS) attacks have emerged as a popular means of causing mass targeted service disruptions, often for extended periods of time. The relative ease and low costs of launching such attacks, supplemented by the current inadequate sate of any viable defense mechanism, have made them one of the top threats to the Internet community today. Since the increasing popularity of web-based applications has led to several critical services being provided over the Internet, it is imperative to monitor the network traffic so as to prevent malicious attackers from depleting the resources of the network and denying services to legitimate users. This paper first presents a brief discussion on some of the important types of DDoS attacks that currently exist and some existing mechanisms to combat these attacks. It then points out the major drawbacks of the currently existing defense mechanisms and proposes a new mechanism for protecting a web-server against a DDoS attack. In the proposed mechanism, incoming traffic to the server is continuously monitored and any abnormal rise in the inbound traffic is immediately detected. The detection algorithm is based on a statistical analysis of the inbound traffic on the server and a robust hypothesis testing framework. While the detection process is on, the sessions from the legitimate sources are not disrupted and the load on the server is restored to the normal level by blocking the traffic from the attacking sources. To cater to different scenarios, the detection algorithm has various modules with varying level of computational and memory overheads for their execution. While the approximate modules are fast in detection and involve less overhead, they provide lower level of detection accuracy. The accurate modules employ complex detection logic and hence involve more overhead for their execution. However, they have very high detection accuracy. Simulations carried out on the proposed mechanism have produced results that demonstrate effectiveness of the proposed defense mechanism against DDoS attacks.*

## 1. INTRODUCTION

A *denial of service* (DoS) attack is defined as an explicit attempt by a malicious user to consume the resources of a server or a network, thereby preventing legitimate users from availing the services provided by the system. The most common DoS attacks typically involve flooding with a huge volume of traffic and consuming network resources such as bandwidth, buffer space at the routers, CPU time and recovery cycles of the target server. Some of the common DoS attacks are SYN flooding, UDP flooding, DNS-based flooding, ICMP directed broadcast, Ping flood attack, IP fragmentation, and CGI attacks [1]. Based on the number of attacking machines deployed to implement the attack, DoS attacks are classified into two broad categories: (i) a single intruder consumes all the available bandwidth by generating a large number of packets operating from a single machine, or (ii) the distributed case where multiple

International Journal of Network Security & Its Applications (IJNSA), Vol.3, No.2, March 2011

attackers coordinate together to produce the same effect from several machines on the network. The latter is referred to as DDoS attack and owing to its distributed nature, it is very difficult to detect. It is highly important that appropriate defense mechanism should be in place to detect such attacks as quickly as possible.

In this paper, a robust mechanism is proposed to protect a web server from DDoS attack utilizing some easily accessible information in the server. The scheme presented in the paper is an extended version of our earlier work described in [2]. This is done in such a way that it is not possible for an attacker to disable the server host and as soon as the overload on the server disappears, the normal service quality resumes automatically. The detection algorithm has several modules that provide flexibility in deployment. While the approximate detection modules are based on simple statistical analysis of the network traffic and involve very less computational and memory overhead on the server, the accurate detection module is based on a statistical theory of hypothesis testing that has more overhead in its execution. The scheme does not affect traffic from the legitimate clients while the detection of the attack is in progress. This aspect of handling DDoS attacks is not taken into account in many of the commercial solutions [3].

The rest of the paper is organized as follows: Section 2 presents some classic DDoS attack types. Section 3 briefly discusses some of the existing work in the literature on defense against DoS attacks. Section 4 presents some salient characteristics of the network traffic, as their understanding is important for design of any defense mechanism for DDoS attacks. Section 5 describes the components of the proposed security system and the algorithms for detection and prevention of attacks. Section 6 presents the simulation results and the sensitivity analysis of the parameters of the algorithms. Section 7 concludes the paper while highlighting some future scope of work.

## 2. DISTRIBUTED DENIAL OF SERVICE ATTACKS

There are two major types of DDoS attacks [4]. The attacks of the first types attempt to consume the resources of the victim host. Generally the victim is a web server or proxy connected to the Internet. When the traffic load becomes very high, the victim host starts dropping packets both from the legitimate users and attack sources. The victim also sends messages to all the sources to reduce their sending rates. The legitimate sources slow down their rates while the attack sources still maintain or increase their sending rates. Eventually, the victim host's resources, such as CPU cycles and memory space get exhausted and the victim is unable to service its legitimate clients. The attacks of the second type target network bandwidth. If the malicious traffics in the network are able to dominate the communication links, then traffics from the legitimate sources are affected. The effects of bandwidth DDoS attacks are usually more severe than the resource consumption attacks. In this section, some classic bandwidth attacks are discussed.

The *SYN flood* attack exploits a vulnerability of the TCP three-way handshake, namely, that a server needs to allocate a large data structure for any incoming SYN packet regardless of its authenticity. During SYN flood attacks, the attacker sends SYN packets with source IP addresses that do not exist or not in use. During the three-way handshake, when the server puts the request information into the memory stack, it will wait for the confirmation from the client that sends the request. While the request is waiting to be confirmed, it will remain in the memory stack. Since the source IP addresses used in SYN flood attacks may be spurious, the server will not receive confirmation packets for requests created by the SYN flood attack. Each half-open connection will remain on the memory stack until it times out. This causes the memory stack getting full. Hence, no request, including legitimate requests, can be processed and the services of the system are disabled. SYN floods remain one of the most powerful flooding methods.

The *smurf* attack is a type of ICMP flood, where attackers use ICMP echo request packets directed to IP broadcast addresses from remote locations to generate denial of service attacks. There are three entities in these attacks: the attacker, the intermediary, and the victim. First, the attacker sends one ICMP echo request packet to the network broadcast address and the request is forwarded to all the hosts within the intermediary network. Second, all of the hosts within the intermediary network send the ICMP echo replies to flood the victim. Solutions to the smurf attack include disabling the IP-directed broadcast service at the intermediary network. Nowadays, smurf attacks are quite rare in the Internet since defending against such attacks are not difficult.

An *HHTP flood* refers to an attack that bombards web servers with HTTP requests. HTTP flood is a common feature in most botnet software. To send an HTTP request, a valid TCP connection has to be established, which requires a genuine IP address. Attackers can achieve this by using a bot's IP address. Moreover, attackers can craft the HTTP requests in different ways in order to either maximize the attack power or avoid detection. For example, an attacker can instruct the botnet to send HTTP requests to download a large file from the target. The target then has to read the file from the hard disk, store it in memory, load it into packets and then send the packets back to the botnet. Hence, a simple HTTP request can incur significant resource consumption in the CPU, memory, input/output devices, and outbound Internet link.

Another important DDoS attack is the *SIP flood* attack. A widely supported open standard for call setup in the *voice over IP* (VoIP) is the *session initiation protocol* (SIP) [5]. Generally, SIP proxy servers require public Internet access in order to accept call setup requests from any VoIP client. Moreover, to achieve scalability, SIP is typically implemented on top of UDP in order to be stateless. In one attack scenario, the attacker can flood the SIP proxy with many SIP INVITE packets that have spoofed source IP addresses [6]. To avoid any anti-spoofing mechanisms, the attackers can also launch the flood from a botnet using non-spoofed source IP addresses. There are two categories of victims in this attack scenario. The first types of victims are the SIP proxy servers. Not only will their server resources be depleted by processing the SIP INVITE packets, but their network capacity will also be consumed by the SIP INVITE flood. In either case, the SIP proxy server will be unable to provide VoIP service. The second types of victims are the call receivers. They will be overwhelmed by the forged VoIP calls, and will become nearly impossible to reach by the legitimate callers.

Figure 1 illustrates another type of bandwidth attack called a *distributed reflector denial of service* (DRDoS) attack, which aims to obscure the sources of attack traffic by using third parties (routers or web servers) to relay attack traffic to the victim. These innocent third parties are also called the *reflectors*. Any machine that replies to an incoming packet can become a potential reflector. The DRDoS attack consists of three stages. The first stage is a typical DDoS attack where the attackers send a large number of packets to the victim host. However, in the second stage, after the attacker has gained control of a certain number of 'zombies' instead of instructing the 'zombies' to send attack traffic to the victims directly, the 'zombies' are ordered to send to the third parties spoofed traffic with the victim's IP address as the source IP address. In the third stage, the third stage, the third parties will then send the reply traffic to the victim, which constitutes a DDoS attack. In comparison to a traditional DDoS attack, the traffic from a DRDoS attack is further dispersed by using the third parties. This makes the attack traffic even more distributed and hence more difficult to identify. Moreover, the source IP addresses of the attack traffic are from innocent third parties. This makes attack source traceback extremely

International Journal of Network Security & Its Applications (IJNSA), Vol.3, No.2, March 2011

difficult. Finally, as noticed in [7], DRDoS attacks have the ability to amplify the attack traffic, which makes the attack even more potent.

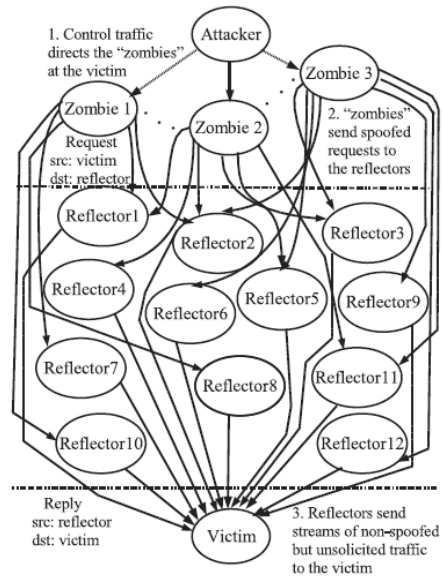

Figure 1. The structure of a distributed reflector denial of service (DRDoS) attack [7]

A particularly effective form of reflector attack is the *DNS amplification attack*. The role of domain name system (DNS) is to provide a distributed infrastructure to store and associate different types of resource records (RR) with Internet domain names. One of the important functions of DNS is to translate domain names into IP addresses. A recursive DNS server accepts a query and resolves a given domain name on behalf of the requester. Generally, a recursive name server will contact other authoritative names servers if necessary and eventually return the query response back to the requester [8]. The sizes of the DNS query response are disproportional. Normally, a query response includes the original query and the answer, which means the query response packet is always larger than the query packet. Moreover, one query response can contain multiple types of RR, and some types of RR can be very large.

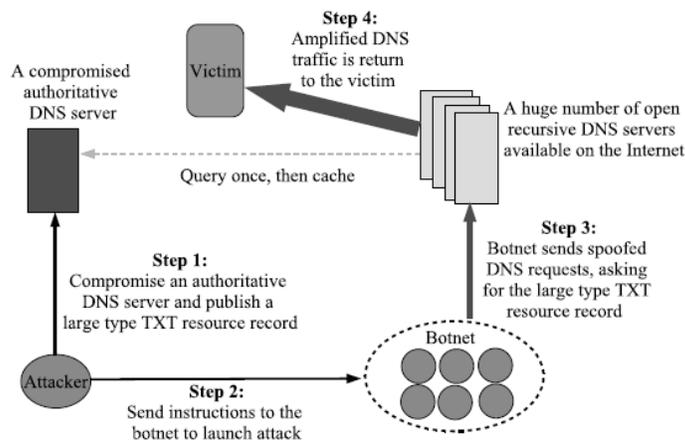

Figure 2. An example of a DNS amplification attack [4]

Figure 2 illustrates an example of a DNS amplification attack that was observed in early 2006 [9]. In this attack scenario, an attacker first compromises an authoritative DNS server and publishes a large type TXT RR. Then the attacker instructs the botnet to send spoofed DNS requests with the victim's IP address to open DNS recursive servers, asking for the large TXT RR. Finally, the open DNS recursive servers resolve the query and return the amplified DNS responses back to the victim. In theory, 140 Mb/s initiating traffic from a botnet can result in a 10 Gb/s DNS flood to the victim. This gives the attacker an opportunity to generate a powerful DDoS attack from even a small botnet. Unfortunately, the opportunity to launch such an attack is widely available in the Internet. According to a survey conducted in 2005 [10], 75% of the 7.5 million external DNS servers allow recursive name service to arbitrary queries. Moreover, attackers do not need to place their own large resource records to implement a successful DNS amplification attack.

## 3. RELATED WORK

Protection against DoS and DDoS attacks highly depends on the model of the network and the type of attack. Several mechanisms have been proposed to solve the problem. However, most of them have weaknesses and fail under certain scenarios. In this section, some of the existing defense mechanisms against DoS and DDoS attacks are discussed briefly.

*Protocol reordering* and *Protocol enhancement* methods make security protocols more robust and less vulnerable to resource consumption attacks [11][12].

*Network ingress filtering* is a mechanism proposed to prevent attacks that use spoofed source addresses [13]. This involves configuring the routers to drop packets that have illegitimate source IP addresses. One of the serious pitfalls of this method is its inability to curtail a flood attack that originates with a spoofed IP address from within the network.

*ICMP traceback* messages are useful to identify the path taken by packets through the Internet [14]. This requires a router to use a very low probability with which traceback messages are sent along with the traffic. Hence, with sufficiently large number of messages, it is possible to determine the route taken by the traffic during an attack. This enables localization of the attacking host.

An approach to overcome the problems associated with ascertaining the validity of IP addresses in ingress filtering is to use the routing information instead of just the source address. *IP traceback* proposes a reliable way to perform hop by hop tracing of a packet to the attacking source from where it originated [15][16]. However, this requires coordinated effort from all the routers in the network along with the path from the victim to the attacker, and examination of the packet logs.

*Deterministic packet marking* (DPM) is another mechanism to detect DoS attacks [17]. It relies on routing information inscribed in the packet header by the routers as the packet traverses the network. This approach leads to an increase in the size of the IP packet header as the size of IP header increases linearly with the number of hops traversed. The resultant variable header size increases the complexity of processing.

*Probabilistic packet marking* (PPM) for IP traceback is a mechanism that attempts to improve DPM [18]. It eliminates IP address spoofing by allowing each router to probabilistically inscribe local path information onto a packet that traverses it [17]. This enables a victim host to localize the attacking source while retaining fixed sized packet headers. The mechanism is dependent on route stability between the attacker and the victim to localize the attacker. A similar mechanism known as *route-based packet filtering* has been proposed in [19], which uses the source and destination addresses on a packet header to ascertain the validity of the route.

Yaar et al. have proposed an approach, called *path identifier* (Pi), in which a path fingerprint is embedded in each packet, enabling a victim to identify packets traversing the same paths through the Internet on a per packet basis, regardless of source IP address spoofing [20]. Pi allows the victim to take a proactive role in defending against a DDoS attack by using the Pi mark to filter out attack packets.

*Pushback* approaches have been proposed to extract attack signatures by rate-limiting the suspicious traffic destined to a congested link [21][22]. Since the DDoS flooding traffic does not follow the end-to-end flow control protocol in the path, it is possible to find the congestion signature using the packet drop statistics. Pushback differentiates attacking traffic from legitimate traffic by monitoring whether the suspicious traffic obeys the end-to-end congestion control.

Gil et al. have proposed a scheme named *MULTOPS* [23] in which routers detect bandwidth attacks using a heuristic based on packet sending rates. Under non-attack circumstances, the packet flow rate in the direction over the Internet is directly proportional to the packet flow rate in the opposite direction. As soon as this condition is violated, an attack is assumed to have occurred. However, efficiency of MULTOPS degrades with randomized IP source addresses.

Mirkovic et al. have proposed a scheme named D-WARD that performs statistical traffic profiling at the edge of the networks to detect new types of DDoS attacks [24]. By monitoring the nominal per-destination type traffic arrival and departure rates of TCP, UDP, ICMP packets, and on observing any abnormal asymmetric behavior of the two-way traffic at the edge router connecting to a stub-network, D-WARD aims at stopping DDoS attacks near their sources.

Zou et al. have presented an adaptive defense system that adjusts its configurations according to the network conditions and attack severity in order to minimize the combined cost introduced by false positives (wrongly identify normal attack as an attack) and false negatives (wrongly identify attack traffic as normal) [25].

*Client side puzzle* and other *pricing algorithms* [26][27][28][29] are effective tools to make protocols less vulnerable to depletion attacks of processing power. However, in case of distributed attacks their effectiveness is debatable.

## 4. NETWORK TRAFFIC CHARACTERISTICS

A proper analysis of the characteristics of network traffic is essential for developing a security system for detection of DDoS attacks. Before presenting the proposed model in Section 5, some characteristics of network traffic are discussed in this Section.

In a wide range of situations, normal network traffic is observed to be consisting of traffic bursts, which are contributed typically by a few high-bandwidth connections [30][31]. On the other hand, attack traffics are characterized by a large number of packets over a small duration of time.

It has been shown that *fair queuing* and its variants are able to isolate individual flow behavior while providing fair service [32]. While these mechanisms may be useful for flow isolation and hence contamination of attacks, they are not suitable for handling millions of flows over the Internet. These mechanisms are thus not practical since they cannot maintain state information for all the routers on a routing path. Caching mechanisms coupled with queue management as in

International Journal of Network Security & Its Applications (IJNSA), Vol.3, No.2, March 2011

LRU-RED have been suggested to solve this problem by maintaining state only for those flows, which consume heavy network resources [30]. These mechanisms typically employ a limited amount of state information and an appropriate state management algorithm to monitor some few dominant flows in the traffic. Thus it is possible to contain these flows (for which state is to be maintained) by appropriate control mechanisms that treat the *cached flows* differently from the rest of the aggregate traffic [30]. For example, fair queuing with two queues (one for cached traffic and the second one for stateless traffic) may be used to limit the cached traffic from taking more than 40% of the link bandwidth.

This poses an intriguing question regarding how to detect and curtail a DDoS attack that involves a large number of attacking hosts. Moreover, an additional requirement for attack detection is the need for facilitating detection of the attack as quickly as possible so as to minimize the extent of damage caused by the attacker on the target server and the network.

Traffic models are very useful resources for identifying the characteristics of network traffic. Several network models proposed in the literature have shown that aggregate network traffic is highly self-similar in nature at all time-scales [33]. This attributes *long-range dependence* (LRD) to the aggregate traffic. However, instantaneous network traffic is highly bursty and does not adhere to the *fractional gaussian* (fGn) *model*. The burstiness results in non-Gaussian marginal distributions, which has been demonstrated using *multifractal wavelet model* [34]. It has been shown that wavelet-based scaling analysis can be used to characterize Internet traffic [35]. Therefore, the scaling properties of wavelets can be effectively tapped to capture the variations in behavior of network traffic during an attack. These are referred to as *wavelet signatures* and are extremely helpful in detection of DoS attacks.

## 5. THE MAJOR SYSTEM COMPONENTS AND ALGORITHMS

This Section describes the traffic model and the attack model on which the proposed security system has been designed. One of the most vital components of the proposed system is known as the *interface* module. Various components of this module are also described in this Section along with the algorithm being used for detection and prevention of attacks.

### 5.1. Traffic Model and Attack Model

In the proposed traffic model *packets* from the network refers to small independent queries to the server (e.g., a small HTTP query or an NTP question-answer). For simplicity, it is assumed that every query causes the same workload on the server. By using the appropriate enhancements (protocol enhancements, crypto hardware, caching etc.) on the server, the workload on the server due different queries can be made very similar. Since the query packets cause workload (memory and processor time consumption) on the server, after a certain time the server cannot handle incoming traffic any further due to memory and processing overloads.

Let us suppose that the attacker uses *A* number of hosts during the attack. When *A* =1, the attack originates from a single source, and when *A* >1, it corresponds to a distributed attack [36]. There are one or more human attackers behind the attacking sources. These attacking sources are machines on the Internet controlled (taken over) by the attacker. It is assumed that the attacking machines use real addresses, and they can establish normal two-way communication with the server, like a host of any legal client. The human attacker hides well behind the attacking machines in the network, which means that after carrying out the attack and after removal of all compromising traces of attack on the occupied machines, there is no way to find a trace leading to him/her.

168

International Journal of Network Security & Its Applications (IJNSA), Vol.3, No.2, March 2011Two types of sources are distinguished: *legal* sources and *attacking* sources. There are *N(t)* legal sources and *A(t)* attacking sources in time slot *t*. In the proposed model, the attacker can reach his/her goal only if the level of attacking traffic is high enough as compared to the level under normal operation. It is assumed that the attacker can control the extra traffic by changing the number of attacking machines and the traffic generated by these machines. It is also assumed that the attacker is powerful and can distribute the total attacking traffic among attacking machines at his/her choice. The reason for using several attacking machines is to make it more difficult for the server to identify and foil them. However, when the attacker uses more machines, it becomes more difficult for him/her to hide the attack. Therefore, the attacker needs to keep the number of attacking hosts at a reasonably small value, i.e., *A(t)* should not be too large. In fact, a trade-off should be made between the ability of the attacker to hide and the efficiency of the attack.

## 5.2. The Interface Module

A DDoS *interface* module is attached to the server at the network side. The interface module may be a software component of the server, a special-purpose hardware in the server host, or an autonomous hardware component attached to the server.

The incoming traffic enters a FIFO buffer. For the purpose of modeling and analysis a *discrete time model* is assumed. Traffic is modeled and processed over unit time slot. The server CPU processes $\mu$ storage units per time slot from the buffer. Since the buffer is fed by a random traffic, there is a non-zero probability of an event of buffer overflow. When a DDoS attack is launched, the incoming traffic quickly increases and the buffer becomes full. At this time, most of the incoming packets will be dropped and the attacker becomes successful in degrading the quality of service of the server. However, the server host will not be completely disabled at this point of time. The goal of the interface module is to effectively identify and disrupt the traffic from the attacking sources so that the normal level of service may be restored promptly.

It is assumed that there are two states of the incoming channel: the *normal state*, and the *attack state*. While in the normal state, there is no DDoS attack on the server, in the attack state, the server is under a distributed attack. Let us assume that the attack begins at time $t^*$, and at time $t^* + \delta$, the interface buffer becomes full. At this time, the TCP modules running at the legal clients and the attacking hosts observe that no (or very few) acknowledgements are being sent back by the server. In order to defend against the DDoS attack, the first task is to detect the point of commencement the attack by making a reliable estimation of the time $t^*$.

Once the time of commencement of the attack is estimated, the next task is to identify the sources of the attack, and to disrupt the traffic arriving from these sources to the server. In the proposed scheme, this identification has been done based on the statistical properties of the traffic flow. The interface module at the server identifies all active traffic sources, measures the traffic generated by these sources, and classifies them into different sets. In order to get reliable measurements of the traffic level, these measurements are carried out during time slots between $t^*$ and $t^* + \delta$. Consequently, the effectiveness of the mechanism is heavily dependent on the time duration $\delta$. During the time $\delta$, the traffic flow between the sources and the server is not affected, i.e., the interface module in the server does not disrupt traffic from the attack sources. It is obviously desirable to have a large value for the time duration $\delta$ so that more time is available for traffic measurement. A large value of $\delta$ can be effectively achieved by using a very large buffer size. It is assumed that the total buffer size *(L)* of the server consists of two parts. The first part $(L_1)$ is designed to serve the normal state of the server. The size of $L_1$ is chosen according to the service rate of the server and the normal probability of packet loss due to the event of a buffer overflow. The size of $L_2$ corresponds to the excess size of the buffer introduced

169

International Journal of Network Security & Its Applications (IJNSA), Vol.3, No.2, March 2011

with the purpose of gaining enough time for traffic measurements during the start-up phase of the attack for identification of the attack sources.

It is assumed that the attack begins at time $t^*$, i.e., all the attacking sources start sending packets at this time onwards. It is also assumed that the network was in normal state at any time $t < t^*$. Let $\hat{t}$ denote the expected value of $t^*$. For the sake of simplicity, it is assumed that the set of active sources is constant during the period of the attack.

Let $T_n(t)$ be the aggregate network traffic from the legal sources (i.e., the normal network traffic), and $T_a(t)$ be the aggregate of the attacking traffic. Let the mean (per time slot) values of the normal and the attack traffic are $\lambda_n$ and $\lambda_a$ respectively.

$$E(T_n(t)) = \lambda_n \quad E(T_a(t)) = \lambda_a \quad (1)$$

Similarly, let the corresponding standard deviations be denoted by $\sigma_n$ and $\sigma_a$. Let $Q$ denote the *apriori* unknown ratio between $\lambda_n$ and $\lambda_a$, i.e. $Q = \lambda_a / \lambda_n$.

As the time of commencement of attack ($t^*$) is earlier than the time of its detection ($\hat{t}$), some precious time is wasted that cannot be used for traffic measurements. To minimize, this loss, the aggregate traffic level is estimated continuously by using a *sliding window* technique. The interface module in the server handles two *sliding time windows*. The longer window has a capacity of $w_l$ slots, and the shorter one has a capacity of $w_s$ slots. In this way, both an *extended-time average* level $\bar{\lambda}(t)$ and a *short-time average* level $\hat{\lambda}(t)$ of the incoming aggregate traffic per slot at time slot $t$.

### 5.3. Algorithms in the Interface Module

The interface module in the server executes four algorithms in order to identify the DDoS attack and the attacking sources. These four algorithms are executed sequentially in the same order as they are mentioned. The algorithms are: (i) algorithm for detection of an attack, (ii) algorithm for identification of the attack sources, (iii) algorithm for disrupting the traffic arriving from the attack sources, and (iv) algorithm for testing whether the attack traffic has been successfully disrupted. In the following subsections, these four algorithms are described in detail.

### 5.3.1. Algorithm for Attack Detection

In order to ensure high availability of the server, an early detection of an attack is of prime importance. As discussed in Section 5.2, the beginning of an attack is assumed to take place at time $\hat{t}$. An approximate determination of $\hat{t}$ can be done in any of the following two ways:

(i) $\hat{t}$ is the point of time when the buffer $L_l$ becomes full.
(ii) $\hat{t}$ is the point of time when the following inequality holds:

$$\hat{\lambda}(\hat{t}) > (1+r)\bar{\lambda}(\hat{t}) \quad (2)$$

In the inequality (2), $r > 0$ is a design parameter. It represents the maximum value of the fraction by which the short-term average of traffic level may exceed the long-term average without causing any alarm for attack on the server. In Section 6, the comparative analysis of the effectiveness of the two approaches in detecting a distributed attack is presented with simulation results.

However, for a more accurate and reliable identification of an attack, a statistical approach based of hypothesis testing is also proposed. In this approach, a large sample of packet arrival pattern on the server is taken for a long duration. The *packet arrival rate* (PAR) at each sample duration is periodically measured and the sample mean ($\bar{X}$) and the sample standard deviation ($\hat{S}$) of the PAR are computed. Let $X_1, X_2, \ldots X_N$ be a sample of $N$ measurement. Then, ($\bar{X}$) and ($\hat{S}$) are given by (3) and (4):

$$\bar{X} = \frac{\sum_{i=1}^{N} X_i}{N} \tag{3}$$

$$\hat{S} = \sqrt{\frac{\sum_{i=1}^{N}(X_i - \bar{X})^2}{N-1}} \tag{4}$$

After the computation of $\bar{X}$ and $\hat{S}$, one-sample *Kolmogorov-Smirnov* (*K-S*) test is applied to test if the samples come from a population with a normal distribution. It is found that *P*- values for all *K-S* tests are greater than $\alpha = .05$. Therefore, it is concluded that the PAR follows a normal distribution. In other words, $\bar{X}$ is normally distributed with an unknown mean, say, $\mu$. The standard value of $\bar{X}$ is computed as in (5):

$$Z = \frac{\bar{X} - \mu}{\hat{S}/\sqrt{N}} \tag{5}$$

In (5), $Z$ is a standard normal variable and satisfies (6):

$$P\{-Z_{\alpha/2} \leq \frac{\bar{X} - \mu}{\hat{S}/\sqrt{N}} \leq Z_{\alpha/2}\} = 1 - \alpha \tag{6}$$

In (6) $\alpha$ is the level of confidence which satisfies $0 \leq \alpha \leq 1$. Equation (6) tells the fact that there is a probability of 1- $\alpha$ of selecting a sample for which the confidence interval will contain true value of $\mu$. $Z_{\alpha/2}$ is the upper 100 $\alpha/2$ percentage point of the standard normal distribution. Therefore, the 100(1- $\alpha$)% confidence interval of $\mu$ is given by (7):

$$\bar{X} - Z_{\alpha/2}\frac{\hat{S}}{\sqrt{N}} \leq \mu \leq \bar{X} + Z_{\alpha/2}\frac{\hat{S}}{\sqrt{N}} \tag{7}$$

The confidence interval in (7) gives both a lower and an upper confidence boundary for $\mu$.

To detect an attack scenario, a threshold value called *maximum packet arrival rate* (MPAR) is defined which distinguishes the normal PAR and the high PAR in an attack. In order to find MPAR, the upper confidence bounds for $\mu$ in equation (7) are obtained by setting the lower confidence bound to -$\infty$ and replacing $Z_{\alpha/2}$ by $Z_\alpha$. A 100(1- $\alpha$)% upper confidence bound for $\mu$ is then obtained from equation (8). The value of $\alpha$ in (8) is 0.025.

$$\mu \leq T_x = \overline{X} + Z_\alpha \frac{\hat{S}}{\sqrt{N}} \tag{8}$$

A robust statistical *t*-test is now applied to verify the difference between normal PAR and the attack PAR. Let $\mu_1$ and $\mu_2$ denote the population means of two traffic flows. The *t*-test is applied to determine the significance of the difference between the two means, i.e. ($\mu_1 - \mu_2$). Let the difference between the two sample means be ($\overline{X}_1 - \overline{X}_2$), and the standard deviation of the sampling distribution of differences is $\sqrt{(\frac{S_1^2}{N_1} + \frac{S_2^2}{N_2})}$. The *t*-statistic is computed in (9).

$$t = \frac{\overline{X}_1 - \overline{X}_2}{\sqrt{(\frac{S_1^2}{N_1} + \frac{S_2^2}{N_2})}} \tag{9}$$

Since the two groups may contain different sample sizes, a weighted variance estimate *t*-test is used. The weighted variance is computed in (10):

$$\hat{S}^2 = \frac{(N_1 - 1)S_1^2 + (N_2 - 1)S_2^2}{N_1 + N_2 - 2} \tag{10}$$

The resultant *t*-statistic is computed in (11):

$$t = \frac{\overline{X}_1 - \overline{X}_2}{\sqrt{\frac{\hat{S}_1^2}{N_1} + \frac{\hat{S}_2^2}{N_2}}} \tag{11}$$

To detect attack traffic, the following hypotheses are tested. The null hypothesis $H_0$: $\mu_1 = \mu_2$ is tested against the alternative hypothesis $H_1$: $\mu_1 \neq \mu_2$. Levene's test is used to assess $H_0$. If the resulting *P*-values of Levene's test is less than a critical value (0.05 in this case), $H_0$ is rejected and it is concluded that there is a difference in the variances of the populations. This indicates that the current traffic flow is an attack traffic. As will be evident in Section 6.3, this accurate statistical algorithm has 100% detection accuracy in all simulation runs conducted on the system. However, it due to complex computational overhead, it is a bit slower than the approximate algorithms.

### 5.3.2. Algorithm for Identification of Attack Sources

It is essential to disrupt the traffic emanating from the attack sources at the interface module of the server after an attack is detected. For this purpose, the interface module must be able to distinguish between the traffic from the attack sources and the normal traffic from legitimate client hosts. As mentioned in Section 4, the distinguishing characteristic of the attack sources is the higher mean ($\lambda_a$) of their aggregate traffic level. It is assumed that the interface module can measure the traffic characteristics of all the active sources at each time instance by recognizing their network addresses.

Starting at time $\hat{t}$, the traffic level corresponding to every source is measured. If an attack was correctly identified, i.e. $t^* < \hat{t} < t^* + \delta$, traffic measurement and analysis can be made over the period $(t^* + \delta - \hat{t})$. Let the aggregate level of traffic be $\hat{\lambda}_r (t^* + \delta)$, and the traffic for the source $i$ be $\hat{\lambda}(i) (t^* + \delta)$. As the exact traffic from the legal sources during the attack cannot be determined, the expression $\overline{\lambda}(\hat{t} - c)$, $(c > 0)$, is used as an estimate of mean aggregate traffic level of the legal sources in time interval $[t^*, t^* + \delta]$, and an estimate for the mean aggregate traffic level of the attacking sources ($\overline{\lambda}_a$) is derived as in (12):

$$\overline{\lambda}_a = \hat{\lambda}_r(t^*+\delta) - \overline{\lambda}(\hat{t} - c) \qquad (12)$$

The set $Z$ of active sources is decomposed into two mutually disjoint sets $Z_n$ and $Z_a$, where the former is the set of *legal sources* and the latter is the set of *attacking sources*. The sets $Z$, $Z_n$ and $Z_a$ will satisfy (13):

$$Z = Z_n \cup Z_a \quad Z_n \cap Z_a = \phi \qquad (13)$$

The identification algorithm produces as output a set $Z_a^*$, which is a subset of the set $Z$ and very closely resembles the set $Z_a$. The closer the sets $Z_a$ and $Z_a^*$ are, the more accurate is the detection of the sources of attacks. The identification of the attacking sources is made by the following two ways:

(i) In this approach, the maximal subset of $Z_a^* = \{i_1, i_2, \ldots, i_L\}$ of $Z$ is computed that corresponds to sources with the highest measured traffic levels so that the inequality (14) is satisfied. The set $Z_a^*$ contains the attack sources.

$$\sum_{j=1}^{v} \hat{\lambda}^{(i_j)}(t^*+\delta) \leq \hat{\lambda}_a \qquad (14)$$

The basis principle for this method is that the attacker always tries to hide himself/herself, and therefore limits the number of attacking sources ($A(t)$). At the same time, to make the attack effective, the attacker intends to send a high volume of attack traffic to the server. Thus, there is a trade-off with the volume of the attack and the number of attack sources. As a result of this trade-off, the volume of traffic emanating from the attacking sources is higher than the volume of traffic from the legitimate client hosts. This criterion is used for identification of attack sources in equation (14).

(ii) In this method, the sources from the set of traffic sources $Z$ which are active during the interval ($\hat{t} - c$), $c > 0$, are omitted and equation (14) is used to identify the attack sources.

### 5.3.3. Algorithm for Disruption of Attack Traffic

Once the attacking sources are correctly identified, the disruption of the traffic emanating from the attack sources is a relatively straightforward task. This is proposed to be done in the following way.

All the incoming packets with source addresses belonging to set $Z_a^*$ are discarded. A filter rule is used at the server inbound interface to discard any incoming packets from the identified sources. Next, any previously stored packet already existing in the interface buffer from these source addresses are deleted to ensure that the server does not process any request from the attack sources.

### 5.3.4. Algorithm for Checking the Success of Attack Traffic Disruption

If the algorithm presented in Section 5.3.3 for traffic disruption from the attack sources executes successfully, then the available buffer size should come back to the level of $L_1$ within a timeout interval, $t\_out$. If this does not happen, additional packets from active sources are to be discarded. The active source that has the highest level of measured traffic is chosen for packet discard, and the available buffer size is checked. If the buffer size is still not restored, another source is chosen for discarding its packets. These steps are repeated until the occupied buffer size comes to the level of $L_1$. The equation (15) gives a conservative estimate for the timeout interval $t\_out$.

$$t\_out = d \cdot \frac{L_2}{\mu - \bar{\lambda}(\hat{t} - c)} \qquad (15)$$

## 6. SIMULATIONS AND RESULTS

In this section, the details of simulations and the results are discussed. The simulation process involved four components as shown in Figure 3. The *interface simulator module* performs the buffering and scheduling of the incoming packets for further processing. It also collects statistical data on traffic for detection of possible attacks, identification of the sources of such attacks, and disrupting communication from such sources.

The simulation program has been written in C and the program is run on a workstation with Red Hat Linux version 9 operating system. A MySQL database is used for storing data related to traffic. The time interval is set at $10^{-6}$ seconds. Statistical data are collected in every second interval. The simulation is done with first 100 seconds as the normal traffic. The attack simulation is started at the $100^{th}$ second and is allowed to continue till the $200^{th}$ second. The simulation is ended with another 100 seconds of normal traffic to test efficacy of the recovery function of the system.

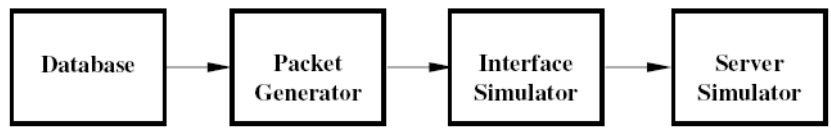

Figure 3. The schematic flow diagram of attack detection simulation process

### 6.1. Simulation Parameters

The traffic arrivals at the interface module are modeled as Poisson process. The packets are stored in a buffer and are passed on to the CPU for further processing by the interface module. The queue type is assumed to be M/M/1. The inter-arrival time and service times are assumed to follow the negative exponential distribution. The number of sources is kept constant throughput the simulation duration.

Poisson model of traffic arrival is chosen as it is particularly suitable for dealing with some Internet protocols if its parameters are set appropriately. *Internet control message protocol* (ICMP), *network time protocol* (NTP) and *domain name service* (DNS) clients send many small packets of constant size with uniformly distributed inter-packet arrival time. These protocols resemble very closely to the assumptions that have been made in the simulation. This makes the results in simulation realistic.

174

International Journal of Network Security & Its Applications (IJNSA), Vol.3, No.2, March 2011Since in practical scenarios, the number of legitimate clients that connect to a server may also vary over a broad range, the following cases are considered:

*Case 1*: For a small corporate server, the number of legal clients is low, say $N(t) = 5$. Assuming that the capacity of the server is high, the average load on the server will be less. Therefore, the number of attacking hosts should be high, say $A(t) = 40$. Hence, in this scenario, for an effective attack we must have $N(t) << A(t)$.

*Case 2*: For a server of medium size, it may be assumed that $N(t) = 50$ and a successful attacker can launch his/her attack from a fewer number of hosts. Thus it may be assumed that $A(t) = 50$ in this case. As the number of legal clients and the number of attacking sources are of comparable size, it is easier for the attacker to hide his/her attack in this case. Therefore, in this situation, $N(t) \approx A(t)$.

*Case 3*: For a global portal server, there can be a very large number of legal clients, say $N(t) = 10000$. In this situation, it is not possible for that attacker to easily estimate the required number of attacking hosts. In this case, it is assumed that the attacker chooses a reasonably high value of $A(t)$, say $A(t) = 5000$, and opts for a very high attacking rate: $\lambda_a = \lambda_n*10$. Therefore, in this case: $N(t) > A(t)$.

In the first simulation, a large number of hosts are taken to test the effectiveness of the proposed mechanism on a large system. The simulation parameters are listed in Table 1.

Table 1. Simulation parameters for Simulation I

| Parameter | Value |
| --- | --- |
| Number of legal clients ($N(t)$) | 10000 |
| Number of attacking hosts ($A(t)$) | 5000 |
| Mean normal traffic rate ($\lambda_n$) | 0.1 |
| Mean attack traffic rate ($\lambda_a$) | 0.4 |
| Service rate ($\mu$) (packets/sec) | 1500 |

With 10000 legal clients and $\lambda_n = 0.1$, the capacity of the server should be at least 1000. However, the attack is successful only when the service rate ($\mu$) is less than 3000 ($\lambda_a*A(t) + \lambda_n*N(t)$). The value of $\mu$ is, therefore, taken as 1500.

The buffer size for normal situation is taken as 40 packets i.e., $L_1 = 40$ (packets). For choosing the size of $L_2$, it is observed that the normal traffic rate is 1000 packets/sec. Thus a safe value of $L_2 = 3000$ (packets) is taken. The values of the parameters of the attack detection algorithm are given in Table 2.

Table 2. Parameters of the attack detection algorithm

| Parameter | Value |
| --- | --- |
| Sliding window size ($w_s$) | 10 sec |
| Tolerance for traffic jump ($r$) | 0.6 |
| Time frame for last correct value of $\lambda$ | 45 sec |

The available time for traffic analysis depends on the value of $\delta$. Therefore, an accurate estimation of the value of this parameter is crucial for effective working of the proposed mechanism. In the simulation work, a constant value ($\hat{\delta} \leq \delta$) for this parameter is used for traffic analysis. It is assumed that the total traffic (normal and attack) is known and its value is $T_n + T_a = 3000$. As the service rate ($\mu$) is 1500, one can expect the buffer $L_1$ to be full after $40/(3000-1500) \approx 0.3$ seconds. The whole buffer ($L= L_1 + L_2$) will be full in $30040/(3000-1500) \approx 200$ seconds. Therefore, a safe estimation of $\hat{\delta} = 10$ is made. In real world situation, one does not have any preliminary knowledge about the attack and so $\delta$ should be estimated over a period of time. For the sake of simplicity, the value of $\hat{\delta}$ is set equal to $w_s$. The algorithm presented in Section 5.3.2 is used for identification of the attacker.

Table 3. Results of Simulation I

| Observed metrics | $\hat{\delta}$ ($\hat{\delta} = w_s$) | | | | |
|---|---|---|---|---|---|
| | 5 | **10** | 20 | 30 | 40 |
| Correctly identified attackers | 2982 | **3784** | 4529 | 4784 | 4892 |
| Filtered legal clients | 1 | **557** | 260 | 132 | 59 |
| Dropped packets | 0 | **0** | 0 | 14251 | 28765 |
| Max. buffer level and corresponding time frame | 29717 (200 s) | **14941 (110s)** | 29732 (119s) | 30040 (120s) | 30040 (120s) |
| Time to restore (after $t^*$) | 149 | **104** | 73 | 71 | 81 |

## 6.2. Simulation I

Table 3 shows the results of the simulation with different values of the window size ($w_s$). It is clear that a larger window size and hence a large $\delta$ gives a more accurate identification of attacks. However, with a larger window size the system is more likely to enter into a situation of buffer overflow. However, during the attack, the buffer will allow for traffic measurement during the initial 20 seconds. After the buffer overflow, the detection algorithm will produce very inaccurate and unreliable results. Therefore it is not worthwhile to increase the window size beyond a limit. On the other hand, as evident from Table 3, when the time window is too short, the algorithm can detect only a very small proportion of the attacking hosts. The determination of an optimum window size is a challenging research problem. In summary, the simulation results In Table 3 show that the mechanism can successfully detect an attack with a window size of 10 seconds.

## 6.3. Simulation II

In this case, a smaller system is simulated with parameters are listed in Table 4. The buffers $L_1$ and $L_2$ are chosen as 40 and 160 respectively. The value of $\delta$ is set equal to $w_s$, i.e. $\hat{\delta} = w_s = 10$. The remaining parameters are kept the same as in simulation I.

Table 4. Results of Simulation I

| Parameter | Value |
|---|---|
| Number of legal clients ($N(t)$) | 50 |
| Number of attacking hosts ($A(t)$) | 50 |
| Mean normal traffic rate ($\lambda_n$) | 0.1 |
| Mean attack traffic rate ($\lambda_a$) | 0.2 |
| Service rate ($\mu$) (packets/sec) | 8 |

In simulation II, experiments are repeated on 500 different sets of input data to have an insight into the statistical properties of the system under normal and attack situations. With different data sets, it is observed that the approximate algorithm (ii) in Section 5.3.1 was faster in detecting the attack in 454 cases. In 42 cases, the attack was correctly identified by both algorithms (i) and (ii) in Section 5.3.1. The accurate detection algorithm presented in Section 5.3.1 could detect all the 50 of attack sources without any filtering of traffic from the legitimate clients in all the 500 simulation runs. Therefore, in terms of detection accuracy and reduced false positives the accurate statistical algorithm for detection outperformed both the approximate algorithms. However, the approximate algorithm (ii) is found to be faster in detecting the attacks. Table 5 summarizes the simulation results.

Table 5. Results of Simulation II

| Observed metrics | Observed values | | |
|---|---|---|---|
| | Min | Avg | Conf. Int. (95%) |
| Traffic restoration time (after $t^*$) | 49 | 114.732 | 1.942 |
| Packets dropped | 0 | 0.695 | 0.321 |
| Normal user filtered (type II error) | 1 | 7.115 | 0.231 |
| Number of attackers filtered | 21 | 32.413 | 0.235 |
| Attack detection time (after $t^*$) | 0 | 2.95 | 0.09 |

## 7. CONCLUSION

The steady evolution of DDoS attacks as a means for achieving political, economic, and commercial gains, and the relative ease, low costs, and limited accountability in launching such attacks, have rendered them one of the top threats to today's Internet services. Although various independent DDoS attack prevention, mitigation, and traceback techniques have been proposed by researchers over the last decades, their relative uptake has been minimal at beast, due to the lack of a robust, fool-proof, and universal DDoS attack defense mechanism. In this paper, a mechanism is presented for detection and prevention of DDoS attacks on a web server. A set of algorithms is presented for attack detection based on traffic analysis and statistical theory of hypothesis testing. To cater to different scenarios, both approximate and accurate detection algorithms are presented which involve varying computational and memory overheads on the server. While the proposed mechanism does not affect the traffic from legitimate clients, it effectively blocks traffic from the attack sources with a very low false positive rate and high detection accuracy. The simulation results demonstrate the effectiveness of the proposed mechanism. Development of an analytical framework for finding an optimum value of the traffic analysis window ($w_s$) and design of a heuristic for faster attack detection with more accuracy are the two areas in which future research work will be carried out.

International Journal of Network Security & Its Applications (IJNSA), Vol.3, No.2, March 2011

179